\documentclass[10pt]{article}

\usepackage[english]{babel}
\usepackage[latin1]{inputenc}
\usepackage{amsmath, amsfonts,amssymb}
\usepackage{amstext}
\usepackage{graphicx}

\usepackage[colorlinks, citecolor=blue, backref]{hyperref}

\newtheorem{theorem}{Theorem}
\newtheorem{proposition}[theorem]{Proposition}
\newtheorem{corollary}[theorem]{Corollary}
\newtheorem{lemma}[theorem]{Lemma}
\newtheorem{definition}[theorem]{Definition}
\newtheorem{remark}[theorem]{Remark}

\newenvironment{proof}{{\bf Proof: }}{\hspace{\fill}$\clubsuit$}

\newcommand{\THEO}[1]{\begin{theorem}{#1}\end{theorem}}
\newcommand{\PROP}[1]{\begin{proposition}{#1}\end{proposition}}
\newcommand{\COR}[1]{\begin{corollary}{#1}\end{corollary}}

\newcommand{\DEF}[1]{\begin{definition}{#1}\end{definition}}
\newcommand{\REMARK}[1]{\begin{remark}{#1}\end{remark}}
\newcommand{\PROOF}[1]{\begin{proof}{#1}\end{proof}}

\def\id{{\mathchoice
     {\rm 1\kern-0.35em{}1}
     {\rm 1\kern-0.35em{}1}
     {\rm 1\kern-0.25em{}1}
     {\rm 1\kern-0.25em{}1}
     }}
\def\be{\begin{equation}}
\def\ee{\end{equation}}

\providecommand{\abs}[1]{\lvert#1\rvert}
\providecommand{\Lie}{\pounds}
\def\<<{``}
\def\R{\mathbb{R}}

\def\D{\partial}

\def\fa{\forall}
\def\map{\longrightarrow}
\def\iff{\Longleftrightarrow}
\def\associa{\longmapsto}

\def\ga{\gamma}
\def\al{\alpha}
\def\B{\beta}
\def\si{\sigma}

\def\ve{\varepsilon}
\def\la{\lambda}
\def\de{\delta}

\def\th{\theta}

\def\/{\, /\;}

\begin{document}
\setlength{\parindent}{0in}

\author{\begin{tabular}{c}
Enrico Bibbona${}^a$, Lorenzo Fatibene${}^{a,b}$, Mauro Francaviglia${}^{a,b,c}$\\
\\
${}^a$ Dipartimento di Matematica, Università degli Studi di Torino (Italy)\\
${}^b$ INFN, Sezione di Torino, Iniziativa Specifica NA12 (Italy)\\
${}^c$ ESG, Università della Calabria (Italy)\end{tabular}}
\title{Chetaev vs. vakonomic prescriptions in constrained field theories with parametrized variational calculus}
\maketitle

\begin{abstract}
Starting from a characterization of admissible Cheataev and vakonomic variations in a field theory with constraints we show how the so called parametrized variational calculus can help to derive the vakonomic and the non-holonomic field equations. We present an example in field theory where the non-holonomic method proved to be unphysical.  
\end{abstract}

\section{Introduction}
At least two different procedures to obtain the field equations for a mechanical problem with non integrable constraints on the velocities have been developed. They are respectively called the vakonomic and the non-holonomic method and are both based on variational principles where a suitable restriction on the set of admissible variations is imposed. In the vakonomic (vak) setting the restriction arises from geometric considerations, while in the non-holonomic (NH) case it is derived from d'Alembert principle.

The question of which one of the two methods produces equations the solutions of which can be physically observed has been extensively studied and it seems (see \cite{exp}) that, at least for a very large class of mechanical constraints, the non-holonomic procedure works better. Nevertheless the vakonomic schema proved to give interesting results in other frameworks, such as optimal control theory (see for example \cite{bloch}).

In field theory, however, the situation is much less clear: both procedures have been generalized to provide field equations and N\"other currents in some cases (see \cite{marsden-art, fgr, en} for vak and \cite{delbinz, deleon, krupkova} for NH), but it is still not evident which one should be better applied in concrete cases. Moreover no fundamental reason justify the NH method since d'Alembert principle cannot be formulated.

Here we aim at contributing to this debate by reformulating both methods in terms of parametrized variational calculus: the use of a parametrization sometime helps to find field equations without the need of additional variables such as Lagrange multipliers.

We also provide few examples. In particular we find that if we interpret matter conservation as a non-integrable constraint in relativistic hydrodynamics,  the non-holonomic methods give non-physical results (every section satisfying the constraint is a solution), while the vakonomic method can be successfully implemented. In our knowledge this is the first field theory example in which one of the two methods has to be rejected, and, surprisingly enough it is exactly the one which works in Mechanics.

\section{Constrained field theories}

Let $C\stackrel{\pi}{\map} M$ be the {\it configuration bundle} whose (global) sections $\Gamma(C)$ represent the fields (by an abuse of language we will often denote bundles with the same label as their total spaces). Let moreover $(x^\mu, y^i)$ be a fibered coordinate system on $C$.

\DEF{A first order {\it Lagrangian} on the configuration bundle 
$C\stackrel{\pi}{\map}M$ ($m= \dim M$) is a fibered morphism
\[ L: J^1 C \map \wedge^m \, T^\ast M.\]}
In local coordinates it can be represented as an horizontal $m$-form $L(x^\mu, y^i, y^i_\mu)\, ds$ on $J^1C$ where $ds$ denote the standard local volume form induced on $M$ by the coordinates $x^\mu$.

\DEF{Let $C\stackrel{\pi}{\map} M$ be the configuration bundle. A {\it constraint} of first order with codimension $r$ is a submanifold $S\subset J^1 C$ of codimension $r$ that projects onto the whole of $C$.}
The constraint be hence expressed by a set of $r$ independent first-order differential equations $\Phi^{(\al)}(x^\mu, y^i, y^i_\mu)=0$.

\DEF{A configuration $\si\in\Gamma(C)$ is said to be {\it admissible} 
with respect
to $S$ if its first jet prolongation lies in S. The space of admissible
configurations with respect to $S$ is
\[ \Gamma_S(C)= \{\si\in\Gamma(C)\/ \rm{Im} (j^1\si) \in S  \}.\] }

\DEF{The set $\{C, L, S\}$ where $C\stackrel{\pi}{\map}M$ is a 
configuration bundle, $L$ is a Lagrangian on $C$ and $S$ is a 
constraint, is called a \<<constrained variational problem''.}

\subsection{Vak-criticality} \label{1}

\DEF{Given a compact submanifold $D\stackrel{cpt}{\subset} M$, a {\it vak-admissible variation} (at first order) of an admissible configuration $\si\in
\Gamma_S(C)$ is a smooth one parameter family of sections $\{\si_t\}_{t\in ]-1,1[}\subset
\Gamma (\pi ^{-1} D)$ such that
\begin{enumerate}
\item $\si_0 =\si|_D$
\item $\fa t \in ]-1,1[,\quad \si_t|_{\D D}= \si|_{\D D} $
\item $\rm{Im}\left(\frac{d}{d t}j^1 \si_t |_{t=0}\right) \in J^1 VC \cap TS$. \label{tre}
\end{enumerate} }

In order to check condition $3$, one has to verify that the vertical vector field $V=\frac{d}{d t}j^1 \si_t |_{t=0}$ satisfies the following condition:
\begin{equation} \frac{\D \Phi^{(\al)}}{\D y^i} V^i + \frac{\D \Phi^{(\al)}}{\D y^i_\mu} d_\mu V^i=0.\label{tng}\end{equation}

\DEF{We say that an admissible section $\si\in \Gamma_S(C)$ is {\it 
vakonomically critical} (or vak-critical) for the variational problem $\{C, L, S\}$ if
$\fa D\stackrel{\text{\tiny{cpt}}}{\subset}M$ for any vak-admissible variation
$\{\si_t\}_{t\in ]-1,1[}\in
\Gamma (\pi ^{-1} D)$, we have
\[\left.\frac{d}{d t} \int_D L \circ j^1
\si_t \right|_{t =0}=0. \] }

An equivalent infinitesimal condition is the following
\begin{gather*}\fa D\stackrel{\text{\tiny{cpt}}}{\subset}M,\, \fa V\in VC\,
\text{  s. t. } \text{Im} [j^1 (V\circ \si)] \in TS\text{ and } V|_{\D 
D}=0,\\
\int_D\left[\frac{\D \mathcal{L}}{\D y^a} V^a + \frac{\D \mathcal{L}}{\D y^a_\mu} d_\mu V^a\right]\, ds=0. \end{gather*}

\subsection{Chetaev criticality}

\DEF{\label{cht} Given a compact submanifold $D\stackrel{cpt}{\subset} M$ a {\it Chetaev-admissible variation} of an admissible configuration $\si\in \Gamma_S(C)$ is a smooth one parameter family of sections $\{\si_t\}_{t\in ]-1,1[}\subset
\Gamma (\pi ^{-1} D)$ such that
\begin{enumerate}
\item $\si_0 =\si|_D$
\item $\fa t \in ]-1,1[,\quad \si_t|_{\D D}= \si|_{\D D} $
\item \label{ttre} the vertical vector field $\frac{d}{d t} \si_t |_{t=0}=V$ on $\si$ with coordinate expression $V=V^i \D_i$ is such that $\displaystyle\frac{\D \Phi^{(\al)}}{\D y^i_\mu} V^i=0$.
\end{enumerate} }

\DEF{We say that an admissible section $\si\in \Gamma_S(C)$ is {\it 
Chetaev critical} for the variational problem $\{C, L, S\}$ if
$\fa D\stackrel{\text{\tiny{cpt}}}{\subset}M$ for any Chetaev-admissible variation
$\{\si_t\}_{t\in ]-1,1[}\in
\Gamma (\pi ^{-1} D)$ we have
\[\left.\frac{d}{d t} \int_D L \circ j^1
\si_t \right|_{t =0}=0. \] }

An equivalent infinitesimal condition is the following
\begin{gather*}\fa D\stackrel{\text{\tiny{cpt}}}{\subset}M,\, \fa V\in VC\,
\text{  s. t. } \frac{\D \Phi^{(\al)}}{\D y^i_\mu} V^i=0\text{ and } V|_{\D 
D}=0,\\
\int_D\left[\frac{\D \mathcal{L}}{\D y^a} V^a + \frac{\D \mathcal{L}}{\D y^a_\mu} d_\mu V^a\right]\, ds=0. \end{gather*}

\subsection{Integrable constraints}

\DEF{Let $C$ and $B$ two bundles on the same base $M$ and let $(x^\mu,y^i)$ and $(x^\mu,z^A)$ be two fibered coordinate systems on $C$ and $B$ respectively. Given a fibered morphism $f:C\map B$ projecting onto the identity of $M$, with coordinate expression $z^A(x^\mu,y^i)$, its first order jet prolongation $j^1 f$ is a fibered morphism $j^1 f: J^1 C\map J^1 B$ with coordinate expression 
\[z^a_\mu = d_\mu z^a = \frac{\D z^A}{\D x^\mu}+ y^i_\mu \frac{\D z^A}{\D y^i} ,\] where the operator $d_\mu$, called {\it formal derivative}, realizes formally the total derivative with respect to $x^\mu$.}

\THEO{Let $S\in J^1C$ be a set of integrable constraints linear in the derivatives, locally expressed as the zero set of the prolongation $\Phi^{(\al)}_\mu= d_\mu f^{(\al)}(x,y)=0$ of a morphism $f:C\map E$ ($E$ is a vector bundle).
With respect to $S$ any Chetaev-admissible variation is also vak-admissible and viceversa.\label{linear}}

\PROOF{The condition for the vertical vector field $V=\frac{d}{d t} \si_t |_{t=0}$ to be relative to a Chetaev admissible variation is that
\begin{equation} \frac{\D \Phi^{(\al)}_\mu}{\D y^i_\nu}V^i= 	\de^\nu_\mu \frac{\D f^{(\al)}(x,y)}{\D y^i}  V^i=0 \quad \iff\quad \frac{\D f^{(\al)}(x,y)}{\D y^i} V^i=0 \label{Ch}\end{equation}
On the other hand, for vak-admissibility the following condition is needed
\[ \frac{\D \Phi^{(\al)}_\mu}{\D y^i}  V^i + \frac{\D \Phi^{(\al)}_\mu}{\D y^i_\nu} d_\nu V^i=0 \quad \iff\quad d_\mu \frac{\D f^{(\al)}}{\D y^i} V^i + \de^\nu_\mu \frac{\D}{\D y^i} f^{(\al)} d_\nu V^i\]
and this is equivalent to
\begin{equation} d_\mu \left( \frac{\D}{\D y^i} f^{(\al)} V^i \right)=0.\label{vakV}\end{equation}

Now obviously \eqref{Ch} implies \eqref{vakV}, while  if \eqref{vakV} holds then $\frac{\D}{\D y^i} f^{(\al)} V^i$ has to be a constant, and being V vanishing at the boundary, \eqref{Ch} holds too.
}

\COR{For any constrained variational principle $\{C, L, S\}$ with integrable constraint $S$ any vakonomically critical section is Chetaev critical.}

\section{Variational calculus with parametrized variations}
\label{2}

\DEF{
A {\it parametrization} of order $1$ and rank $1$
of the set of constrained variations is a couple
$(E,\mathbb{P})$, where
$E$ is a vector bundle $E\stackrel{\pi_E}{\map} C$, while 
$\mathbb{P}$ is a fibered morphism (section)
\[\mathbb{P}:  J^1 C \map (J^1 E)^\ast\otimes_{J^1C} VC.\]
}
If $(x^\mu, y^a,\varepsilon^A)$ are local fibered coordinates on $E$ and $\{\D_a\}$ is the induced fiberwise natural basis of $VC$, a parameterization of order $1$ and rank $1$ associates to any section $y^a(x)$ of $C$ and any section $\varepsilon^A(x, y)$ the section
\begin{equation} \big[p^a_A \varepsilon^A + p^{a\phantom{A}\mu}_A d_\mu \varepsilon^A\big]\, \D_a \label{parcoor}\end{equation}
of $VC$, where $p^a_A$ and $p^{a\phantom{A}\mu}_A$ are functions of $\big(x^\nu, y^b(x), \D_\nu y^b(x) \big)$, while $\varepsilon^A$ depends on $(x^\nu,y^b(x))$.

\DEF{\label{xxx}Given a compact submanifold $D \subset M$ an {\it
admissible variation} of a configuration $\si\in
\Gamma(C)$ on $D$ is a smooth one parameter family of sections $\{\si_t\}_{t\in ]-1,1[}\subset
\Gamma (\pi ^{-1} D)$ such that
\begin{enumerate}
\item $\si_0 =\si|_D$
\item $\fa t \in ]-1,1[,\quad \si_t|_{\D D}=\si|_{\D D} $
\item \label{333}there exists a section $\ve\in\Gamma(E)$ such that  $\frac{d}{d t}j^1 \si_t |_{t=0} =<\mathbb{P}\,|\,j^1\ve>\circ j^1 \si$ and $(\rho \circ \si)|_{\D D}=0$. 
\end{enumerate} }

\DEF{The set $\{C, L, \mathbb{P}\}$, where $C\stackrel{\pi}{\map}M$ is a 
configuration bundle, $L$ is a Lagrangian on $C$ and $\mathbb{P}$ is a 
parameterization of the set of constrained variations, is called a \<<parametrized variational problem''.}

\DEF{We define {\it critical} for the parametrized variational problem $\{C, L, \mathbb{P}\}$ those sections of $C$ for which, for any compact $D \subset M$ and for any admissible variation $\{\si_t\}$ defined on $D$ one has 
\[\left.\frac{d}{d t } \int_D L \circ j^1
\si_t \right|_{t =0}=0.\]} 

Accordingly, if we use the trivial parametrization $\mathbb{P}: C\map VC^\ast\otimes_C VC$ that to any $p\in C$ associates the identity matrix of $V_pC$ then the third condition becomes empty and we recover free variational calculus.

For an ordinary variational problem with Lagrangian $L= \mathcal{L}\, ds$ criticality of a section of $C$ is equivalent, in local fibered coordinates $(x^\mu, y^a)$, to the fact that for any compact $D\subset M$ and for any $V\in \Gamma(VC)$ such that $j^{k-1} (V\circ\si)|_{\D D}=0$ one has
\[ \int_D\left[\frac{\D \mathcal{L}}{\D y^i} V^i(x,y(x)) +
\frac{\D \mathcal{L}}{\D y^i_\mu}
d_\mu V^i (x,y(x))
\right]\, ds=0. \]

Explicit calculations (see \cite{lo}) show that the above local coordinate expressions glue
together with the neighboring giving rise to the following global one
\begin{gather}\label{iiii} \fa D\stackrel{\text{\tiny{cpt}}}{\subset}M,\, \fa V\in VC\,
\text{  s. t. } (V\circ \si )|_{\D 
D}=0,\notag \\ \int_D<\de L\:|\:
j^1 V>
\circ j^1 \si =0 \end{gather} where $\de L$ is a fibered 
morphism
\begin{equation}
\label{deltaL}\de L: J^1 C \map (J^1 VC)^\ast \otimes_{J^1 C} 
\Lambda^m T^\ast M.
\end{equation}

To define criticality for first order parametrized variational problems we have to restrict variations to those $V\in \Gamma(VC)$ in \eqref{iiii} with $(V\circ \si )|_{\D 
D}=0$ that can be obtained through the parametrization from a section $\varepsilon$ of $E$ satisfying $j^{1} (\varepsilon \circ \si)|_{\D D}=0$.
If  $(x^\mu, y^a, \varepsilon^A)$ are local fibered coordinates on $E$ and $ (p^a_A \varepsilon^A + p^{a\phantom{A}\mu}_A  d_\mu \varepsilon^A)\D_a$ is the local representation of $<\mathbb{P}\,|\,j^1 \varepsilon>\circ j^1 \si$, criticality holds if and only if for any compact $D\subset M$ for any section $\varepsilon$ with coordinate expression $\varepsilon^A(x,y)$ such that both $\varepsilon^A(x,y(x))=0$ and $d_\mu \varepsilon^A(x,y(x))=0$ for all $x\in \D D$, we have
\begin{equation} \int_D\left[\frac{\D \mathcal{L}}{\D y^a} (p^a_A \varepsilon^A + p^{a\phantom{A}\mu}_A  d_\mu \varepsilon^A) +
\frac{\D \mathcal{L}}{\D y^a_\mu}
d_\mu (p^a_A \varepsilon^A + p^{a\phantom{A}\nu}_A  d_\nu \varepsilon^A)\right]\, ds=0.\label{local} \end{equation}

To set up a characterization of critical sections in terms of a set of differential equations let us introduce the following procedure: let us split the integrand of \eqref{local} into a first summand that factorizes $\ve^A$ (without any derivative) plus the total derivative of a second term (a general theorem ensures that this splitting is unique; see \cite{en}). To do this, we integrate by parts the derivatives of $\ve$ in the integrand of equation \eqref{local}. What we get is
\[\int_D \big(\mathbb{E}_A \ve^A + d_\mu(\mathbb{F}_A^\mu \ve^A + \mathbb{F}_A^{\mu\nu} d_\nu \ve^A)\big)\,ds=0\]
with
\begin{align*}{}&\mathbb{E}_A = \left(\frac{\D L}{\D y^a} - d_\nu \frac{\D L}{\D
y^a_\nu}\right) p^a_A  - d_\mu \left[ \left(\frac{\D L}{\D y^a} - d_\nu
\frac{\D L}{\D y^a_\nu}\right){p^a_A}^\mu\right] \\
&\mathbb{F}^\mu_A = \left(\frac{\D L}{\D y^a} - d_\nu
\frac{\D L}{\D y^a_\nu}\right){p^a_A}^\mu
+\frac{\D L}{\D y^a_\mu} p^a_A  \\
&\mathbb{F}_A^{\mu\nu}=
\frac{\D L}{\D y^a_\mu}{p^a_A}^\nu .\end{align*}

In \cite{en} we have shown that the coefficients $\mathbb{E}_A$, $\mathbb{F}^\mu_A$ and $\mathbb{F}_A^{\mu\nu}$ are the components of two global morphisms
\[\begin{aligned}{}&\mathbb{E}(L, \mathbb{P}): J^3 C \map 
E^\ast\otimes_C
\Lambda^m T^\ast M\\
&\mathbb{F}(L, \mathbb{P},\ga): J^2 C \map
(J^{1}E)^\ast\otimes_{J^1 C} \Lambda^m T^\ast M \end{aligned}\]
and that to the whole procedure can be given a global meaning in terms of variational morphisms and global operations between them.
The same can also be done for higher order Lagrangians and for 
higher rank and higher order parametrizations.

\subsection{Vak-adapted parameterization}

\DEF{Let $C\stackrel{\pi}{\map}M$ be the configuration bundle.
A {\it parametrization}
\[\mathbb{P}_S:  J^1 C \map (J^1 E)^\ast\otimes_{J^1C} VC\]
of the set of constrained variations is said to be {\it vak-adapted} to the constraint $S\subset J^1C$ if for all $ \varepsilon\in\Gamma(E)$ and $\si\in \Gamma_S(C)$
the vertical vector field
$j^1 (<\mathbb{P}\,|\,j^1 \varepsilon>\circ j^1 \si)$ has image in $TS$.
\label{vakadapt}}

To be vak-adapted to a constraint given by the equations $\Phi^{(\al)}=0$ one has to check that the parametrization with coordinate expression \eqref{parcoor} authomatically implement condition \eqref{tng} or, in formula, that we have

\begin{equation} \frac{\D \Phi^{(\al)}}{\D y^i} \big[p^a_A \varepsilon^A + p^{a\phantom{A}\mu}_A d_\mu \varepsilon^A\big] + \frac{\D \Phi^{(\al)}}{\D y^i_\mu} d_\mu \big[p^a_A \varepsilon^A + p^{a\phantom{A}\mu}_A d_\mu \varepsilon^A\big]=0.\label{tng2}\end{equation}

\DEF{A parametrization $\mathbb{P}_S$ vak-adapted to a constraint $S$ is said to be {\it vak-faithful on $\si\in \Gamma_S(C)$ to $S$} if for all $V\in VC \text{ such 
that both } \rm{Im}[j^1 (V \circ\si) ]\in
TS$ and $ (V \circ\si)|_{\D D}=0$ hold, there exist a section $\varepsilon\in\Gamma(E)$ such that $<\mathbb{P}\,|\,j^1 \varepsilon>\circ j^1 \si = V$ and $j^{1} (\varepsilon\circ\si)|_{\D D}=0$.}

The fundamental problem of the existence of a faithful parametrization that is vakonomically adapted to a constraint $S$ has been studied recently in \cite{lag-mult}, where a universal faithful parameterization has been found for any constraint satisfying certain (quite restrictive) conditions. However we stress that also  non-faithful parameterizations can be useful for some specific tasks (see Section \ref{esempiuzzo} and Remark \ref {faithful?}).

\PROP{Let $\si\in \Gamma_S(C)$ be a vakonomically critical section 
for the constrained variational problem
$\{C, L, S\}$; then for any adapted parametrization $\mathbb{P}_S$ the section $\si$ is $\mathbb{P}_S$-critical.}
\PROOF{We have
\begin{gather*}
\si\in \Gamma_S(C) \text{ is vak-critical}\\
\Updownarrow \\
\fa D\stackrel{\text{\tiny{cpt}}}{\subset}M,\; \fa \text{ adm. var. }
\{\si_\ve\},\quad\left.\displaystyle\frac{d}{d \ve } \int_D L \circ
j^1
\si_\ve
\right|_{\ve =0}=0\\
\Updownarrow \\
\fa D\stackrel{\text{\tiny{cpt}}}{\subset}M,\, \fa V\in VC\,
\text{  s. t. } \rm{Im} [j^1 (V \circ \si)]\in TS \text{ and }\si|_{\D D}=0,\\
\int_D<\de L\:|\:
j^1 V>
\circ j^1 \si =0\\
\phantom{(a)}\Downarrow \text{(a)}\\
\displaystyle\fa D\stackrel{\text{\tiny{cpt}}}{\subset}M,\; \fa \varepsilon \in
\Gamma(E)\text{ s. t. }j^{2}\varepsilon|_{\D D}=0 ,\\
\int_D \Big< <\de
L\,|\,j^1 \mathbb{P}'>\,\Big|\, j^{2}\varepsilon\Big> \circ 
j^{3} \si =0\\
\Updownarrow \\
\mathbb{E}\circ j^{3}\si=0.
\end{gather*}

Step (a) is not an equivalence since there can be admissible 
infinitesimal variations
vanishing at the boundary with their derivatives up to the desired 
order that do not come
from sections of the bundle of parameters that do vanish on the 
boundary. The last
equivalence holds in force of Stokes's theorem, the vanishing of
$j^{l+k-1}\varepsilon$ on the boundary and
the independence of the generators of $E$. }
\COR{Let $\si\in \Gamma_S(C)$ be an admissible $\mathbb{P}_S$-critical section for 
the constrained variational problem $\{C, L, S\}$ and let 
also $\mathbb{P}_S$ be
faithful to $S$ on $\si$; then $\si$ is vakonomically critical for the 
constrained variational
problem $\{C, L, S\}$.}

\subsection{Chetaev-adapted parameterization}

\DEF{Let $C\stackrel{\pi}{\map}M$ be the configuration bundle.
A {\it parametrization}
\[\mathbb{P}_S:  J^1 C \map (J^1 E)^\ast\otimes_{J^1C} VC\]
of the set of constrained variations is said to be {\it Chetaev-adapted} to the constraint $S\subset J^1C$ if for all $ \varepsilon\in\Gamma(E)$ and $\si\in \Gamma_S(C)$
the vertical vector field
$<\mathbb{P}\,|\,j^1 \varepsilon>\circ j^1 \si$ satisfies condition \ref{ttre} of Definition \ref{cht}.
}

In coordinates, given the expression \eqref{parcoor} for the parametrization the condition reads as
\begin{equation}\frac{\D \Phi^{(\al)}}{\D y^a_\nu} [p^a_A \varepsilon^A + p^{a\phantom{A}\mu}_A d_\mu \varepsilon^A]=0 \label{chet}\end{equation}

\DEF{A parametrization $\mathbb{P}_S$ Chetaev-adapted to a constraint $S$ is said to be {\it Chetaev-faithful on $\si\in \Gamma_S(C)$ to $S$} if for all $V\in VC$ such 
that both $\frac{\D \Phi^{(\al)}}{\D y^i_\mu} (V\circ \si)^i=0$ and $(V \circ\si)|_{\D D}=0$ hold there exist a section $\varepsilon\in\Gamma(E)$ such that $<\mathbb{P}\,|\,j^1 \varepsilon>\circ j^1 \si = V$ and $j^{1} (\varepsilon\circ\si)|_{\D D}=0$.}

As we did in the previous Section we can prove the following proposition:

\PROP{Let $\si\in \Gamma_S(C)$ be a Chetaev critical section 
for the constrained variational problem $\{C, L, S\}$. For any Chetaev-adapted parametrization $\mathbb{P}_S$ the section $\si$ is also $\mathbb{P}_S$-critical.\label{rapporto}}

\COR{Let $\si\in \Gamma_S(C)$ be an admissible $\mathbb{P}_S$-critical section for 
the constrained variational problem $\{C, L, S\}$ with respect to a faithful $\mathbb{P}_S$; then $\si$ is also vakonomically critical.}

\section{Examples}

In literature very few examples of Lagrangian field theories with constraints are present and the question whether the Chetaev or the vakonomic rule produces equations whose solutions are physically observed is still open. Let us present here two examples: the first is a classic in Mechanics with non-holonomic constraint, while the second, to our knowledge is the first example of a field theory where the vakonomic method seems to be preferable to the non-holonomic one (the opposite as in Mechanics).

\subsection{A skate on an inclined plane}
\begin{figure}[htb]
\caption{A skate on an inclined plane}
\label{fig1}
\begin{center}
\fbox{\includegraphics{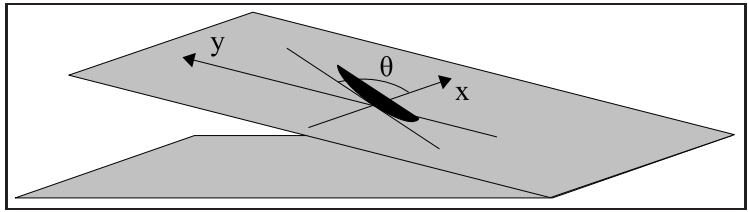}}
\end{center}
\end{figure}

This is the model of a skate (or better a knife edge, as called in \cite{bloch}) that moves on an inclined plane keeping the velocity of its middle point (that is also the unique point of contact with the plane, allowing for rotations) parallel to the blade (see figure \ref{fig1}). 

The kinematic variables are the coordinates $(x,y)$ of the contact point and the direction $\th$ of the blade (thus the configuration bundle $C$ is $\R\times\R^2\times S^1\map \R$).
The Lagrangian is 
\[ L=\frac{m}{2}(\dot{x}^2 + \dot{y}^2) + \frac{I}{2}\dot{\th}^2 -  m g_{\text{eff}} y \]
while the constraint $S\subset J^1 C$ is given by
\[\dot{x}\sin\theta - \dot{y}\,\cos\theta=0.\]

\subsubsection{The non-holonomic setting}

A Chetaev admissible variation for this constraint is a vector field $V \in \Gamma (VC)$, locally identified by the three components $(V_x, V_y, V_{\th})$ with respect to the natural base $(\frac{\D}{\D x},\frac{\D}{\D y},\frac{\D}{\D \th})$, that satisfies condition \ref{ttre} of Definition \ref{cht}: 
\[\sin\theta V_x - \cos\theta V_y=0.\]
A parametrization of admissible variations can be found by solving the previous equation as follows.

Let us introduce the subbundle $K\subset VC$ identified as the kernel of the vector bundle morphism $(V_x, V_y, V_\th)\associa V_x-V_y$. It has a two dimensional fiber and using in the fiber coordinates $(W_1, W_2)$ the fibered immersion in $C$ reads as $(W_1, W_2)\associa(W_1,W_1, W_2)$.
In formal language a faithful Chetaev-adapted parameterization (with zero order and zero rank) is the fibered morphism
\begin{gather*}\mathbb{P}_S: C \map V^\ast B\otimes_C VC \/ \fa \si\in\Gamma (C), \fa W \in \Gamma (VB)\\
<\mathbb{P}_S\,|\,(W_1,W_2)>\circ\, (x(t), y(t), \th(t)) = (\cos \th W_1, \sin\th W_1, W_2).\end{gather*}
Varying the Lagrangian along this parameterization one get the following first variation formula:
\begin{gather*}\big<\de L\,|\, j^1<\mathbb{P}_S\,|\,W>\big>\circ\, j^1\si=\\
=m\dot{x}( \cos \th W_1 )^\cdot -mg_{\text{eff}}(\sin\th W_1) + m\dot{y}( \sin\th W_1 )^\cdot + I\dot{\th}\dot{W_2}=\\
= - m[\ddot{x}\cos\th + (g_{\text{eff}} + \ddot{y})\sin\th]\, W_1 - I \ddot{\th} W_2 +\left[ m (\dot{x} \cos\th + \dot{y} \sin\th) W_1 + I\dot{\th}W_2 \right]^\cdot
\end{gather*}
so that the equations of motion are:
\[\left\{\begin{aligned}{}&\ddot{x}\cos\th + (g_{\text{eff}} + \ddot{y})\sin\th=0\\
&\ddot{\th}=0\end{aligned}\right.\]
\REMARK{One can check that these equations are exactly the same that can be derived from the traditional rule $\mathbb{E}_i= \la_{(\al)}\frac{\D \Phi^{(\al)}}{\D \dot{y}^i}$, where $\mathbb{E}_i$ are the Lagrange equations of the unconstrained variational problem (see {\rm\cite{bloch, garcia-vak}}.}

\subsubsection{The vakonomic setting}

A vak admissible variation for this constraint is a vector field $V \in \Gamma (VC)$, locally identified by the three components $(V_x, V_y, V_{\th})$ with respect to the natural base $(\frac{\D}{\D x},\frac{\D}{\D y},\frac{\D}{\D \th})$, that satisfies (see Definition \ref{vakadapt}) the following condition: 
\[\sin\theta \dot{V_x} - \cos\theta \dot{V_y} + (\cos\th\dot{x}+\sin\th \dot{y})V_\th =0.\]
A parametrization of admissible variations can be found by solving the previous equation.
Let us introduce the vector subbundle $K\subset VC$ identified by $V_\th=0$.
In formal language a faithful vak-adapted parameterization (with order and rank both equal to $1$) if $\dot{y}\cos\th+\dot{x}\sin\th\neq 0$ is the fibered morphism
\begin{gather*}\mathbb{P}_S: J^1C \map (J^1 VB)^\ast\otimes_{J^1C} VC \/ \fa \si\in\Gamma (C), \fa V \in \Gamma (VB)\\
<\mathbb{P}_S\,|\,j^1(V_x,V_y)>\circ\, (x(t), y(t), \th(t)) = \left(V_x,V_y,\frac{\sin\th \dot{V_x} -\cos\th\dot{V_y}}{\dot{y}\cos\th+\dot{x}\sin\th}\right).\end{gather*}
Varying the Lagrangian along the parameterization gives the following first variation formula:
\begin{gather*}\big<\de L\,|\, j^1<\mathbb{P}_S\,|\,W>\big>\circ\, j^1\si=\\
=m\dot{x} \dot{V_x} -mg_{\text{eff}}V_y + m\dot{y}\dot{V_y} + I\dot{\th}\left( \frac{\sin\th \dot{V_x} -\cos\th\dot{V_y}}{\dot{y}\cos\th+\dot{x}\sin\th} \right)^\cdot.\\
\end{gather*}
Integration by parts of the derivatives of the variations leads to the following equations of motion:
\begin{equation}\left\{\begin{aligned}{}&m\ddot{x}=I\left( \frac{\ddot{\th}\sin\th}{\dot{y}\cos\th + \dot{x}\sin\th } \right)^\cdot\\
&m\ddot{y} + m g_{\text{eff}}=-I\left( \frac{\ddot{\th}\cos\th}{\dot{y}\cos\th + \dot{x}\sin\th } \right)^\cdot \end{aligned}\right.\label{zioni}\end{equation}
\REMARK{One can check that these equations are exactly the same that can be derived from the Lagrange multiplier traditional rule (see {\rm\cite{bloch, garcia-vak}}), in fact from the variation of the Lagrangian $L'=L+\la_{(\al)}\Phi^{(\al)}$ one gets the following equations
\[
\left\{\begin{aligned}
{}&\dot{x}\sin\theta - \dot{y}\,\cos\theta=0\\
&m\ddot{y} = (\la \sin\th)^\cdot\\
&m\ddot{y} +mg_{\text{eff}}= -(\la \cos\th)^\cdot\\
&I\ddot{\th}=  \la( \cos\th \dot{x}+ \sin\th\dot{y})
\end{aligned}\right.\]
and if $\dot{y}\cos\th+\dot{x}\sin\th\neq 0$ one can solve the last one for $\la$ and substitute in the others to get again equations \eqref{zioni}.
}

\REMARK{Let us notice that due to the particularly simple form of the constraint equation we can solve it for $\th$ in some open subset of the domain getting $\th=\rm{arctan}\frac{\dot{y}}{\dot{x}}$. This also is the fundamental reason for which it is so easy in this case to find a vak-adapted parameterization. If one now substitutes this expression into the Lagrangian, reducing the configuration bundle but increasing the order of the Lagrangian, and then one varies it with respect to the independent variable $(x,y)$ one gets again equations \eqref{zioni}.\label{rema}}

\subsubsection{Comparison}
For a comparison of the solution we defer the reader to \cite{bloch}, where it has been shown that for some given trajectories the vakonomic and non-holonomic are very much different. For an experimental test of the real observability of vak and NH trajectories in a different mechanical system has been carried on in \cite{exp} where the authors found that realistic trajectories obey NH equations.

\subsection{Relativistic hydrodynamics}

Here we present a field theory example where the vakonomic method and the non-holonomic one give very different result. In particular we find that the non-holonomic theory seems to be non physical (every admissible section is a solution).
Let us consider a region $M$ of spacetime with metric $g_{\al\B}$ (here we consider it to be fixed, but if we want to study the coupling with gravity the formalism can do it as well, see \cite{en}) filled with a barotropic and isoentropic fluid. The world lines of the fluid particles and its density describe completely the configuration of the system.
There are different methods to describe the kinematics. We choose to use a vector density $J^\mu ds_\mu$ physically interpreted as follows: let the unit vector field $u^\mu (g_{\al\B},J^\mu) = \frac{J^\mu}{\abs{J}}$ be tangent to the flow lines in every point, and the scalar $\rho(g_{\al\B},J^\mu) =\frac{\abs{J}}{\sqrt{\abs{g}}}$ be the density of the fluid.
The configuration bundle $C$ is thence that of vector densities of weight $-1$ (the transformation rule of $J^\mu$ for a coordinate change $x^{\mu'} =x^{\mu'}(x^\mu )$ with Jacobian matrix $M$ is $J^{\mu'}= M^{\mu'}_\mu J^\mu \det M^{-1}$).
The dynamics of the system is ruled by the Lagrangian 
\[L =-\sqrt{\abs{g}}\: [\rho\;(1+ e (\rho))]\, ds\]
where the scalar $e(\rho)$ is the internal energy of the fluid from which we can derive the pressure $P= \rho^2 \frac{\D e}{\D \rho}$.
The vector density $J^\mu$ cannot take arbitrary values because of the continuity equation of the fluid 
\[\D_\mu J^\mu=0\]
that needs to hold. It acts as a non integrable constraint $S\subset J^1C$ on the derivatives of the fields.

Some further details on this physical system can be found in \cite{h-e, brown, kijowski-art, taub}.

\subsection{The vakonomic setting}
\label{esempiuzzo}
A vak-admissible variation of the field $J$ with respect to the constraint $S$ is given by a vector field $V\in \Gamma (VC)$ represented in local coordinates as $V^\mu \frac{\D}{\D J^\mu}$, whose first jet is tangent to $S$, that is to say $\D_\mu V^\mu=0$.

A vak-admissible parameterization is given by the morphism
\[\mathbb{P}_S: J^1 C \map (J^1 TM)^\ast \otimes_{J^1C} VC\]
such that $\fa \si \in \Gamma(C)$ and $\fa X \in \Gamma(TM)$
\[<\mathbb{P}_s\,|\,j^1 X>\,\circ j^1\si= \Lie_X J^\mu \frac{\D}{\D J^\mu}= (d_\nu
J^\mu X^\nu - J^\nu d_\nu X^\mu + J^\mu d_\nu X^\nu)\frac{\D}{\D J^\mu}\]

Varying the Lagrangian along the given parameterization gives the following expression
\[<\de L\,|\,j^1 X>\, \circ \si= -\sqrt{\abs{g}}\frac{\D\mu}{\D\rho}\frac{J^\cdot_\mu}{\rho} (d_\nu
J^\mu X^\nu - J^\nu d_\nu X^\mu + J^\mu d_\nu X^\nu)\, ds \]
where $\mu=\rho\;(1+ e (\rho))$ and the following identities hold
\[ \rho \, \frac{\D \mu}{\D \rho} = \mu +P
\hbox{\qquad and \qquad} 
\rho \, d_\nu \left( \frac{\D \mu}{\D \rho} \right) = d_\nu P. \]
Integrating by parts the derivatives of $X$ one finds the following field equations
\begin{equation}(u^\cdot_\mu u^\nu + \de^\nu_\mu)  \nabla_\nu P + (\mu + P )\: u^\nu \nabla_\nu u^\cdot_\mu =0.\label{flu}\end{equation}

\REMARK{In literature one can also find a different description of the system where the fundamental fields are three scalars $R^a (x^\mu)$ physically interpreted as the labels identifying the \<<abstract fluid particle'' passing throw the point $x^\mu\in M$. The quantities $J^\mu$, $u^\mu$ and $\rho$ are then defined as suitable functions of the fundamental fields and their first derivatives that automatically solve the constraint. This description is dynamically equivalent to our one and it performs, as in Remark \ref{rema}, a reduction of the configuration bundle (three fundamental fields instead of four) increasing the order of the Lagrangian in such a way that the variations of the fundamental fields automatically preserve the constraint. We defer the reader to the following references {\rm \cite{taub, brown, kijowski-art}} for further details.}

\begin{figure}[hbt]
\caption{Non faithfulness of $\mathbb{P}_S$}
\label{fig3}
\begin{center}
\fbox{\includegraphics{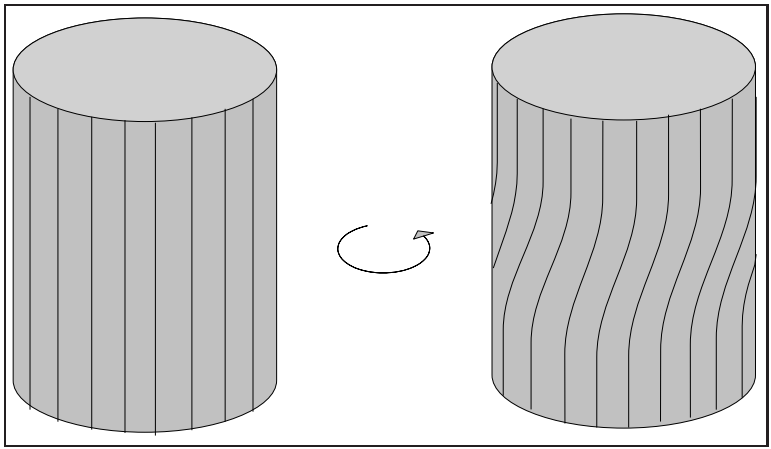}}
\end{center}
\end{figure}

\REMARK{\label{faithful?}The parameterization we used is vak adapted to the constraint $\D_\mu J^\mu=0$, but not faithful to it in general. Varying $J^\mu$ along $\mathbb{P}_S$ is equivalent to drag the flow lines along a vector field $X$ vanishing on the boundary of integration and adjusting $\rho$ so that the constraint is preserved (see \cite{h-e}, section 3.3, example 4). Depending on the solution sometimes it is possible to find variations of $J^\mu$ tangent to $S$ that cannot be produced by means of $\mathbb{P}_S$ and a vector field $X$ vanishing on the boundary of integration. An example is drawn in figure \ref{fig3}: the twisting do not affect $J^\mu$ on the boundary of the cylinder, nevertheless it is generated by a vector field $X$ which does not vanish on the upper part of the boundary itself!
In formal terms we have that $\Lie_X J^\mu|_{\D D}=0$ also if $X|_{\D D} \neq 0$. 
Nevertheless, this is exactly what we want to do from the physical viewpoint: when we think to the congruence of curves that represent our fluid and we imagine to vary them leaving the boundary fixed, we mean that the curves are fixed not only their tangent vectors!
To support our choice to vary fields along our parametrization we stress that also in the alternative approach of \<<abstract fluid particles'' variations leave unchanged the particle identification on the boundary.
Solutions of the Euler-Lagrange equations \eqref{flu} are not necessarily vak-critical solutions of the variational problem with the constraint $S$ given by $\D_\mu J^\mu=0$, but anyway they represent the motions of physical fluids.}

\subsubsection{The non-holonomic setting}

A Chetaev-admissible variation of the field $J$ with respect to the constraint $\Phi=\D_\mu J^\mu=0$ is given by a vector field $V\in \Gamma (VC)$ represented in local coordinates as $V^\mu \frac{\D}{\D J^\mu}$ that satisfies condition \ref{ttre} of Definition \ref{cht}:
\begin{equation}\frac{\D \Phi}{\D \D_\al J^\mu} V^\mu=0\qquad\iff\qquad V=0.
\end{equation}
For the relativistic fluid, thus, the unique Chetaev-admissible variation is identically vanishing.
One cannot define any non-trivial variational framework with an empty set of variations and insisting on this route one obtain that any section $\si\in \Gamma(C)$ is Chetaev-critical and thanks to proposition \ref{rapporto} also $\mathbb{P}_S$-critical for every Chetaev adapted parameterization.
This conclusion is clearly non-physical.

\section{Conclusions}
We have shown how the parametrized variational calculus can contribute to the study of non-holonomic and vakonomic field theories. We have defined the notion of vak-criticality and Chetaev-criticality and compared them with the one of criticality with respect to a vak-adapted or a Chetaev adapted parametrization.
We have also shown examples in Mechanics and Field Theory. In particular we think that relativistic hydrodynamics is the first case where it has been shown that the vakonomic method is preferable to the non-holonomic one (the opposite result with respect to mechanics!).
We still cannot guess why this occurs, nor whether it is a general rule for field theories; nevertheless it seems to us that it is an interesting occurrence and it deserve to be further investigated.

\bibliographystyle{alpha}
\bibliography{bibliografia}
\end{document}